\documentclass[prb,amsmath,amssymb,eqsecnum]{revtex4}
\usepackage{graphicx}
\usepackage{amsmath}
\usepackage{amssymb}
\usepackage{euscript}
\usepackage{epsfig}
\usepackage{rotating}
\usepackage{color}
\usepackage{hhline}
\usepackage[ps2pdf]{hyperref}

\def\be{\begin{equation}}
\def\ee{\end{equation}}
\def\bfl{\begin{flushleft}}
\def\efl{\end{flushleft}}
\def\bfr{\begin{flushright}}
\def\efr{\end{flushright}}
\def\bc{\begin{center}}
\def\ec{\end{center}}
\def\ben{\begin{enumerate}}
\def\een{\end{enumerate}}
\def\bit{\begin{itemize}}
\def\eit{\end{itemize}}

\def\dzn{,\kern-0.1em,}
\def\d#1{{#1\kern-0.4em\char"16\kern-0.1em}}
\def\D#1{{\raise0.2ex\hbox{-}\kern-0.4em 31}}
\begin{document}

\title{Chaotic fluctuation of temperature on environmental interface exchanging energy
by visible and infrared radiation, convection and conduction}

\author{D.T.Mihailovi\'c \footnote{e-mail: guto@polj.ns.ac.yu}}
\affiliation{Faculty of Agriculture, University of Novi Sad, Trg
D. Obradovi\'ca 8, 21 000 Novi Sad, Serbia}
\author{D.Kapor}
\affiliation{Department of Physics, Faculty of Sciences,
University of Novi Sad,  Trg D. Obradovi\'ca 4, 21 000 Novi Sad,
Serbia}
\author{M.Budin\v cevi\'c}
\affiliation{Department of Mathematics, Faculty of Sciences,
University of Novi Sad,  Trg D. Obradovi\'ca 4, 21 000 Novi Sad,
Serbia}
\date{\today}

\begin{abstract}

The concept of environmental interface is defined and analyzed
from the point of view of the possible source of non-standard
behaviour. The energy balance equation is written for the
interface where all kinds of energy transfer occur. It is shown
that under certain conditions, the discrete version of the
equation for the temperature time rate turns in to the well-known
logistic equation and the conditions for chaotic behaviour are
studied. They are determined by the Lyapunov exponent. The
realistic situation when the coefficients of the equation vary
with time, is studied for the Earth-environment general system.

\end{abstract}


\maketitle
%------------------------------------------------------------------
\section{Introduction}
%------------------------------------------------------------------

The concept of  "balance",  either global or local in any given
context, it is undoubtedly the cornerstone in the increasing
number of environmental problems. The question is: Why are the
environmental problems in the focus now? One particular answer can
be found in a hierarchy of the main scientific problems in this
century. According to the physicists, the world scientific
community, will be occupied in this century among others, by the
environmental problems that are primarily expressed through the
problem of climate changes [1, 2] as well as many other linked
problems that comprises a wide range of time and spatial scales.
This is the first time in the history of science that the
environmental problems take the place at the research front of the
sciences. The question, why it is happening now and why it will go
on happening in the future, could be answered by the well known
fact that in the scientific as well as in other worlds the main
"dramatic event" takes place at the interface between either two
media  or two states [3]. The field of environmental sciences is
abundant with various interfaces and it is the right place for
application of new fundamental approaches leading towards better
understanding of environmental phenomena. We define the
environmental interface as an interface between the two, either
abiotic or biotic environments each, which are in a relative
motion exchanging energy through biophysical and chemical
processes and fluctuating temporally and spatially regardless of
its space and time scale. In our opinion, this definition broadly
covers the unavoidable multidisciplinary approach in environmental
sciences and also includes the traditional approaches in sciences
that are dealing with the environmental space less complex than
any one met in reality. The wealth and complexity of processes at
this interface determine that the scientists, as it often seems,
are more interested in a possibility of non-linear "dislocations"
and surprises in the behaviour of the environment than in a smooth
extrapolation of current trends and a use of the approaches close
to the linear physics [3 - 6]. To overcome the current situation
we have to do the following: (a) establish a way in approaching
the non-linear physics and the non-linearity in describing the
phenomena in physics as well as environmental sciences, and (b)
solve or, at least, understand the problem of predictability.
These two problems are problems par excellence of the methodology.
Their successful solving will help us avoid the current problems
in mathematical, physical, biological and chemical interpretation
of the nature. In the next section, we write down the equation for
the energy balance at the environmental interface. We show that
under certain conditions, its discrete version becomes well-known
logistic equation leading to the chaotic behaviour of the
temperature. The conditions for such behaviour are investigated
and particular examples studied. The analysis is based on the
Lyapunov exponent.

\maketitle
%------------------------------------------------------------------
\section{Energy balance equation for environmental interface}
%------------------------------------------------------------------

We see the outside world, i.e. the world of phenomena ({\it
ambience}) from the observer's perspective (its inner world). In
the ambience there are {\it systems} of different levels of
complexity and their environments. System in the ambience is a
collection of precepts while whatever lies outside, like the
component of a set, constitutes the environment [7,8]. The "fate"
of science lies in the fact that it is focused on the system [8].
Furthermore, to be able to anticipate something, we describe the
system  by the states (determined by observations) while the
environment is characterized through its effects on the system.
The environmental interface as a complex system is a suitable area
for occurrence of the irregularities in temporal variation of some
physical or biological quantities describing their interaction.
For example, such interface can be placed between: human or animal
bodies and surrounding air, aquatic species and water and air
around them, natural or artificially built surfaces (vegetation,
ice, snow, barren soil, water, urban communities) and atmosphere
[4-6,9,10], etc. The environmental interface of different media
was recently considered for different purposes [3-6, 11-13]. In
these systems visible radiation provides almost all of the energy
received on the environmental interface. Some of the radiant
energy is reflected back to space. The interface also radiates, in
the thermal waveband, some of the energy received from the sun.
The quantity of the radiant energy remaining on the environmental
interface is the net radiation   (the net radiation energy
available on the surface when all inward and outward streams of
radiation have been considered as seen in Fig. 1 where all kinds
of energy are expressed in terms of the flux density), which
drives certain physical processes important to us. The energy
balance equation may be written  as

\begin{equation} c_{i} \frac{d T_{i}}{d t} = R - H - E - S \end{equation}
where $c_{i}$ is the environmental interface soil  heat capacity
per unit area, $T_{i}$  is the environmental interface temperature
(EI), $t$ is the time,$H$ and $E$ are  the sensible and the latent
heat, respectively,  transferred by convection, and $S$  the heat
transferred by conduction into deeper layers of  underlying
matter.\\

            {\bf Figure 1}\\

Schematic diagram of terms included in the energy balance equation
for environmental interface: (a) net radiation ($R$) that includes
(1) visible radiation, (2) infrared radiation of underlying matter
and (3) infrared counter radiation of the gas; (b) sensible($H$)
and latent heat ($E$) transferred by convection and (c) heat ($S$)
transferred by conduction. \\

The sensible heat is calculated as $C_{H} (T_{i} - T_{a})$  where
$C_{H}$ is the sensible heat transfer coefficient and $T_{a}(t)$
is the gas temperature given as the upper boundary condition. The
heat transferred into underlying soil material is calculated as $
C_{D} (T_{i} - T_{d})$ where $C_{D}$ is the heat conduction
coefficient while $T_d (t)$ is the temperature of deeper layer of
underlying matter that is given as the lower boundary condition.
Following Bhumralkar [14] the net radiation term in Eq. (1) can be
represented as $C_{R} (T_{i} - T_{a})$ where $C_{R}$ is the
radiation coefficient. According to [15], for small differences of
$T_{a}$ and $T_{d}$, the expression for the latent heat can be
written in the form $C_{L}\; f(T_{a})\; [b(T_{i} - T_{a})+ b^{2}
(T_{i} - T_{a})^{2}]/2$. Here $C_{L}$is the latent heat transfer
coefficient, $f(T_{a})$ is the gas vapor pressure at saturation
and $b$ is a constant characteristic for a particular gas.
Calculation of time dependent coefficients , and can be found in
[16]. After collecting all terms in Eq. (1) we get

\begin{equation}  \frac{d T_{i}}{d t} =  A_{1}(T_{i} -
T_{a})- A_{2} (T_{i} - T_{a})^{2} - A_{3} (T_{i} - T_{d})
\end{equation}
where $A_{1} = [C_{R} - C_{H} - C_{L} b f(T_{a})]/c_{i}, \;\;
A_{2} = b^{2} f(T_{a})/(2c_{i})$, and $A_{3} = C_{D}/c_{i}$. This
is a non-linear Riccati type differential equation that
practically always has to be solved numerically, i.e.,
\begin{equation}  {\bf D}T_{i} = {\bf F}_{n} \end{equation}                                                                                                                                        (3)
where ${\bf D}$ is the finite difference operator defined as ${\bf
D}T_{i} = (T_{i, n + 1} - T_{i,n})/{\bf D}t $, $n$ the time level,
${\bf D}t $ is the time step and ${\bf F}_{n}$ is the r.h.s. of
Eq. (2.2) defined at the $n$th time level.

\maketitle
%------------------------------------------------------------------
\section{Numerical results on chaos in temperature on environmental interface}
%------------------------------------------------------------------

We consider Eq. (2.2) with the lower boundary condition, at some
time interval, given in the form
\begin{equation} T_{d} = T_{a} -
(c_{i}/C_{D})\frac{d T_{a}}{d t}\end{equation} expressing slow
temperature changes in both the environment and underlying
material. Then equation becomes

\begin{equation}  \frac{d \xi}{d t} = A_{0} \xi -
A_{2} \xi^{2}  \end{equation}
where $\xi = T_{i} - T{a}$, and
$A_{0} = A_{1} - C_{D}/c_{i}$. Substituting the time derivative by
the finite difference operator ${\bf D}$ in this equation and
after some transformations, we obtain

\begin{equation} \Gamma_{n + 1} = A_{1}^{p} \Gamma_{n} (1 -
\Gamma_{n}) \end{equation} where $\Gamma = (A_{2}^{p}/ A_{1}^{p})
\xi, \;\; A_{1}^{p} = 1 + A_{0} {\bf D} t$ and $A_{2}^{p} = A_{2}
{\bf D}t$. This equation has the same form as the well known
logistic difference equation $\Gamma_{n + 1} = \beta \Gamma_{n} (1
- \Gamma_{n}) $ where $ \beta$ is a constant (see, e.g. [17] ,
among others). Conditions for occurrence of deterministic chaos
for logistic mapping given by this equation are: (a)$0 \leq \Gamma
\leq 1$ and (b)$\; 3.57 \leq \beta \leq 4$. Here we have to bear
in mind that $A_{1}^{p}$ depends on discrete "time" $n$. However,
chaos is still expected, although we can give no quantitative
prediction of the parameter regime. We analyze Eq. (3.3) in the
following way. With ${\bf D}t_{p} = 1/A_{0}$ we indicate the
scaling time range of energy exchange at the environmental
interface including coefficients, that express all kind of energy
reaching and departing the environmental interface. For any chosen
time interval, for solving Eq. (3.3), there always exists ${\bf
D}t_{p,f} = Min[{\bf D}t_{p} (c_{i}, C_{R}, C_{H}, C_{L})]$ when
energy at the environmental interface is exchanged in the fastest
way by radiation, convection and conduction. If we define
dimensionless time $\tau = {\bf D}t/ {\bf D}t_{p,f}$, then Eq.
(3.3) becomes
\begin{equation} \Gamma_{n + 1} = (1 + \tau) \Gamma_{n} (1 -
\Gamma_{n}). \end{equation} Regarding to order of magnitude of
parameters included in coefficients $A_{1}^{p}$ and $A_{2}^{p}$,
the condition (a) is always satisfied. From (b) we get  the
interval ($ \beta = 1 + \tau$ ) where the solution is chaotic,
i.e.,$\; 2.57 \leq \tau \leq 3$. We analyze now the occurrence of
the chaos in solution of Eq. (3.4). Since conditions (a) and (b)
are necessary but not sufficient for identification of the chaos,
we shall calculate values of the Lyapunov exponent  for situations
when these two conditions are satisfied.  Here we define the
Lyapunov exponent $\lambda_{L}\;$ for the case of Eq. (3.4), which
has a single degree of freedom $\Gamma$ which depends on discrete
"time" $j$ following the form [18]
\begin{equation} \lambda_{L} = lim_{n \rightarrow \infty} \frac{1}{n} \sum_{j = 1}^{n} ln
| P'(\Gamma_{j})|  \end{equation} where $ P(\Gamma) = (1 + \tau)\;
\Gamma  (1 - \Gamma)$.

To examine how changes in $A_{0}$ and $A_{2}$ determine the
irregularities in behaviour of the temperature at the
environmental interface obtained from the energy balance equation,
we perform numerical experiment. We solve Eq. (3.2) for different
values of $A_{0}$ and $A_{2}$, varying them in a broad range of
energy exchange coefficients. After that we calculate the
corresponding Lyapunov exponent $\lambda_{L}$  as a function of
$\tau$. Figure 2 depicts dependence of Lyapunov exponent $
\lambda_{L}$ on $\tau$. It takes mostly positive values indicating
chaotic fluctuations of $\Gamma$. However, inside of the chaotic
interval there are a lot opened periodical "windows" where
$\lambda_{L} < 0$. \\

{\bf Figure 2} \\

Dependence of Lyapunov exponent $\lambda_{L}$ on the dimensionless
time $\tau$. \\

Now, we give a specific example of the above analysis applied to
the  interface of the Earth and its environment. It is now
accepted that the Earth is a complex system which consists of the
biota and their environment. These two elements of the system are
closely coupled: the biota regulate the environment (e.g., climate
on planetary scale) and, in turn, the environment restricts the
evolution of the biota and dictates what type of life can exist.
Changes in one part will influence the other, being opposed by
negative feedback or enhanced by positive feedback, and this may
lead to oscillation or chaos in the system.  In the system
considered, the values of the dimensionless time $\tau$  for long
term atmospheric integration are in the interval $1 \leq \tau \leq
3$. Corresponding bifurcation map given in Fig. 3 depicts  the
regions with the chaotic fluctuations of  temperature at the
environmental interface biota-surrounding air. \\

            {\bf Figure 3} \\

Bifurcation diagram of $\Gamma$ as a function of dimensionless
time $\tau$ in the long term atmospheric integration. \\

\maketitle
%------------------------------------------------------------------
\section{Conclusion}
%------------------------------------------------------------------
The main aim of the report was to indicate the possibility of
chaotic behaviour of the temperature at the interface of two media
where the energy is exchanged by all three known mechanisms. The
conditions for such phenomenon are discussed. Once the
dimensionless time is introduced,  one can study the problems on
rather different scales in the same manner (scaling approach).
\\

{\bf Acknowledgement} \\

The research work described in this paper has been funded by the
Serbian Ministry of Science and Environmental Protection under the
project "Modeling and numerical simulations of complex physical
systems ", No. ON141035 for 2006-2010.

%------------------------------------------------------------------

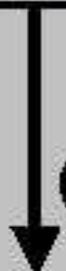
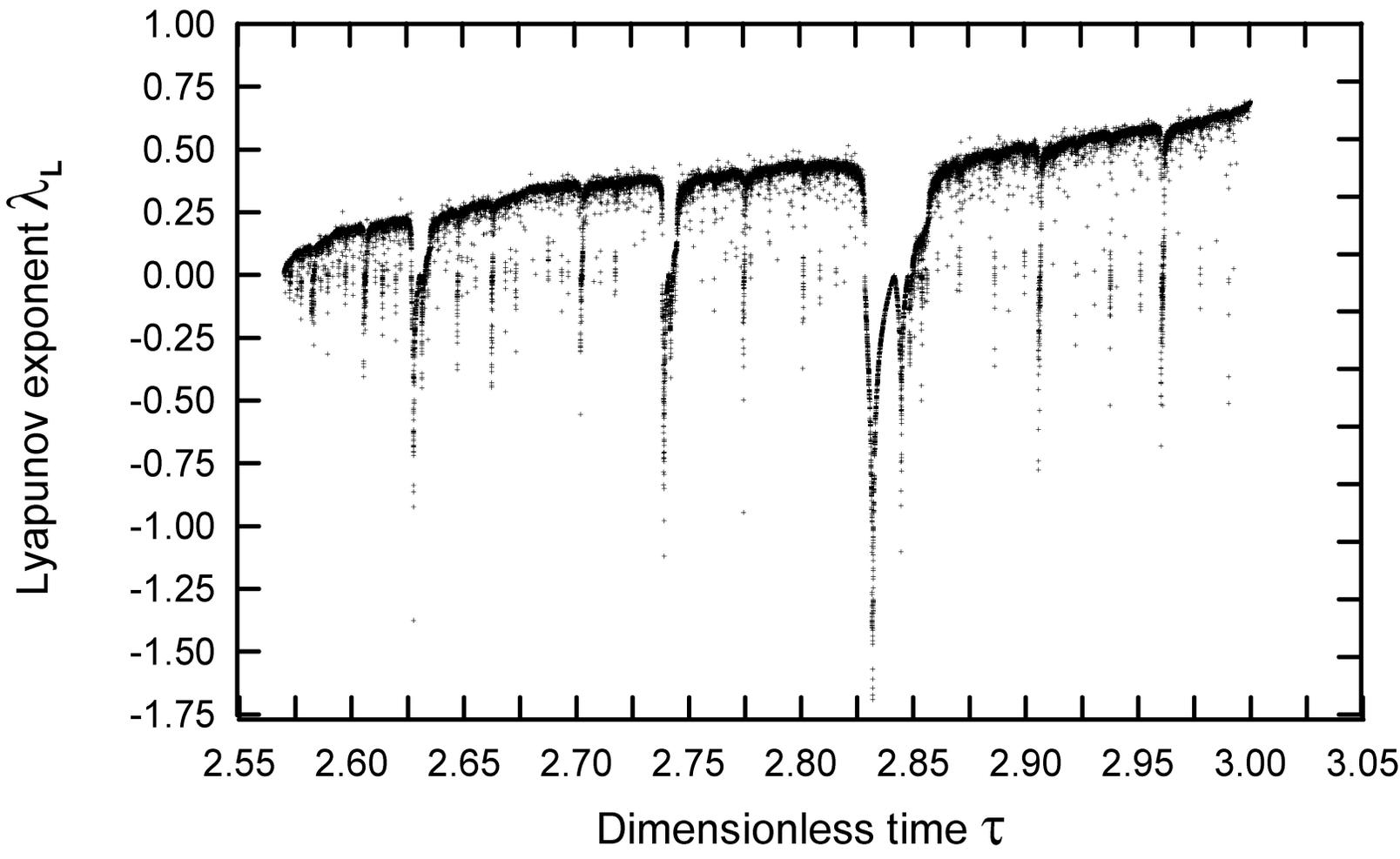
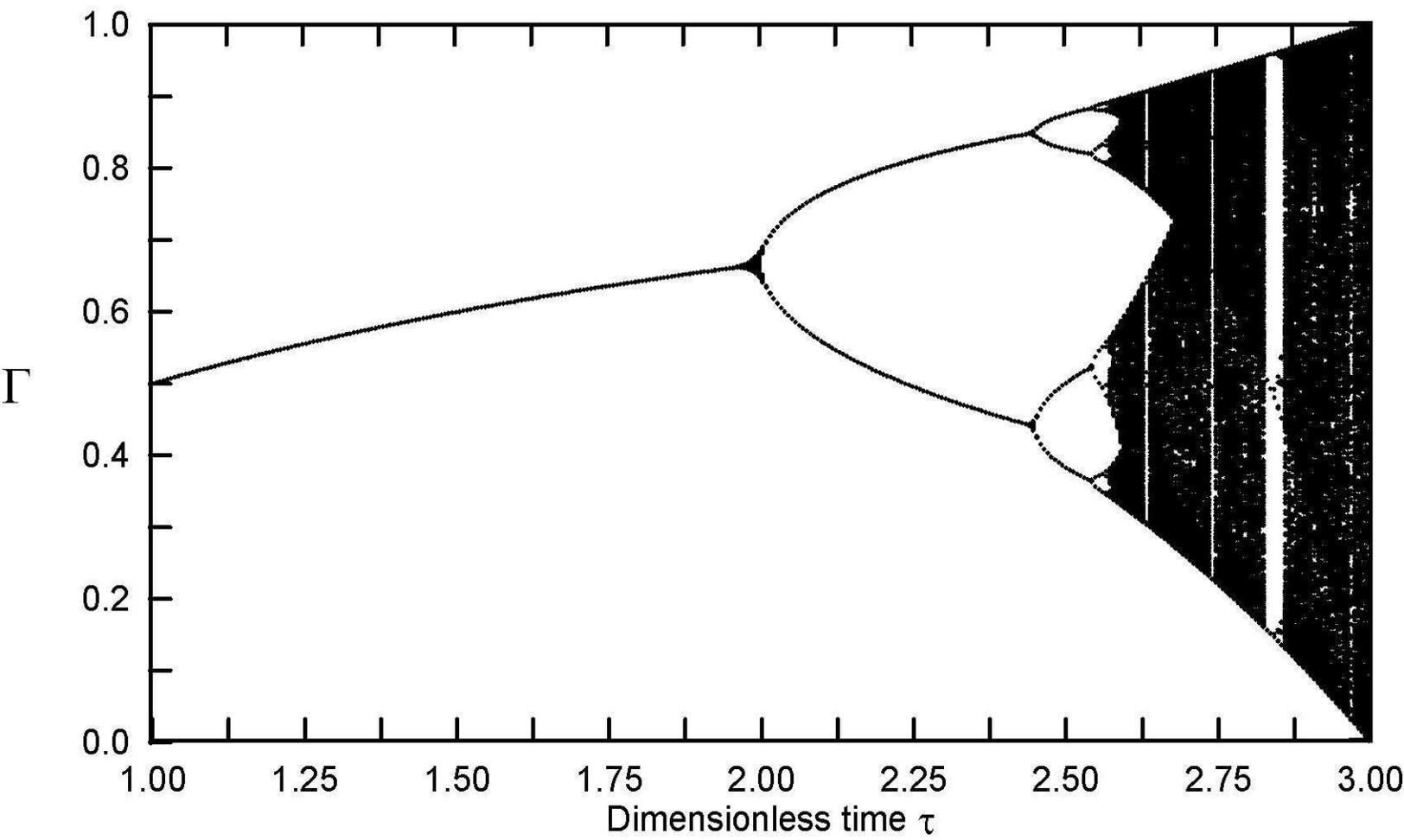
\end{document}